\documentstyle[preprint,aps,pra,epsfig]{revtex}
\tighten
\begin{document}
\draft
\widetext
\preprint{Version: 1.0 }

\title{Proton structure effects in muonic hydrogen}

\author{Krzysztof Pachucki 
\thanks{E-mail address: krp@fuw.edu.pl}}  
\address{
Institute of Theoretical Physics, Warsaw University,
Ho\.{z}a 69, 00-681 Warsaw, Poland.}
\maketitle

\begin{abstract}
The proton structure effects, including finite size, polarizability
and self--energy is considered and their influence on energy levels
of muonic hydrogen is recalculated. A new theoretical prediction
for the Lamb shift is presented together with improved values 
of all known QED contributions. 
\end{abstract}
\pacs{PACS numbers 36.10 Dr, 12.20 Ds, 31.30 Jv}

The precision tests of quantum electrodynamics (QED) in atomic systems
have reached the level, where nuclear structure effects become significant.
Moreover, the lack of accurate data on low energy structure
functions of the nucleus strongly limits theoretical predictions. 
The well known example
is the hydrogen hyperfine structure splitting, where measurement is 
6 orders of magnitude more precise than the current theoretical predictions.
Other example is the Lamb shift in hydrogen, where inaccuracy 
in the proton charge radius dominate other theoretical uncertainties.
Since many years it was the motivation to study the pure QED systems
like muonium and positronium, where strong interaction effects
are negligible or well estimated  at the precision level of interest. 
A new possibilities in the improvement of QED  test on bound systems
appeared with the project of measurement of the Lamb shift
in muonic hydrogen \cite{hau}. 
Being bound in the ground state, muon with its 200
heavier mass compared to the electron,
penetrates the proton and become sensitive to the distribution of charge,
magnetic moment or the polarizability. The main goal of the measurement
of the Lamb shift in muonic hydrogen is a precise determination
of mean square proton charge radius. It will verify older measurements
of proton charge radius based on the  elastic electron--proton scattering.
The results of these experiments had to be extrapolated to very low 
$q^2=0$ momentum transfer, and therefore might be not very reliable.
In contrast the muonic Lamb shift measurement aims to improve the precision
of the proton charge radius by at least 10 times. However it would require the 
calculation of all other contributions with the comparable or better precision.
Few years ago, in the  summary  of all known up to date QED 
effects on muonic hydrogen energy levels \cite{krp2}, 
we concluded that three--loop vacuum polarization have limited the precision
of theoretical predictions by 0.01 meV, and pointed attention to
the proton polarizability effect, which was not calculated that time.
The progress on the experimental side
with the preparation of the measurement of muonic hydrogen Lamb shift
has stimulated further theoretical works. 
The rather difficult three--loop vacuum polarization contribution
has recently been calculated by Kinoshita and Nio \cite{kin} with the result
\begin{equation}
\Delta E(2P-2S) = 0.007\,6\;{\rm meV}\,.
\end{equation}
A new contribution coming from hadronic vacuum polarization has been studied
by Friar and collaborators \cite{fr1}, who obtained the following result
\begin{equation}
\Delta E(2P-2S) = 0.011\,3(3)\;{\rm meV}\,.
\end{equation}
The proton polarizability correction has been analyzed by Rosenfelder in \cite{ros}.
His estimate for this effect is 
\begin{equation}
\Delta E(2P-2S) = 0.017(4)\;{\rm meV}\,.
\end{equation}
Another recent work \cite{fau}  gives a similar result of $\sim 0.018$ meV.
The calculation of proton polarizability effect is affected by the lack
of precise data on low--energy proton structure functions.
In this paper we present  another estimate of this effect
together with the complete review of all other proton structure contributions.
We would like to emphasize the importance of the proton self--energy 
effect and the related problem with the meaning of mean square charge radius.  
In other words we  present the relation of charge radius as obtained
from atomic spectroscopy measurements with that based on the electron 
scattering data.  

It is well known, the shift of atomic energy levels due to
finite charge distribution of atomic nucleus.
It is given by the formula
\begin{equation}
E_{FS} = \frac{2}{3 n^3}\,\alpha^4\,\mu^3\,\langle r^2\rangle\,
\delta_{l0}\,, \label{4}
\end{equation}
which gives $-3.862(108)$ meV contribution to 2P--2S splitting,
were we used  $r=0.862(12)$ fm from \cite{rp}. From this one concludes,
that the measurement of muonic hydrogen with the precision of 0.01 meV
will lead to the tenfold improvement in the proton charge radius.
This formula in (\ref{4}) accounts for most of the proton structure effects. 
Any correction beyond that
is much smaller and usually neglected for light (electronic) atoms. 
It is our aim to review them in the context of muonic hydrogen Lamb shift. 

Since the leading contribution to the 2P-2S splitting
comes from the electron vacuum polarization (e.v.p.), the
second order corrections from combined finite size and e.v.p. effects
are nonnegligible. They have been calculated in \cite{krp1}, 
here we present a little more accurate result
\begin{equation}
\Delta E(2P-2S) = -r^2\,0.0282 = -0.0209(6)\;{\rm meV}\,.
\end{equation}
Further corrections are due to  pure photon exchange terms only.
The $O(m\,\alpha^5)$ correction is given by two--photon scattering 
amplitude with external momenta on mass shell, see Eq. (\ref{7}).
It is the main contribution, which we analyze in this work.
However, lets consider first the small corrections beyond this 
two--photon exchange approximation, namely that of order $m\,\alpha^6$.
Since they are small, it is sufficient to calculate them
in the external field approximation, or in other words in the limit
of infinite nucleus mass. The energy levels are obtained then
from the Dirac equation in the potential of the finite size nucleus.
They have been considered in details by Friar in \cite{fr2}.
In the logarithmic approximation they are given by two terms
\begin{equation}
E = E_{FS}\,\alpha^2\,\ln\alpha\,\biggl(
\frac{2}{3}\,\mu^2\,\langle r^2\rangle-1\biggr) 
\end{equation}
and contribute the amount of $-0.000\,9$ meV to the 2P-2S splitting
and thus are almost negligible. 

The correction given by the two--photon exchange is 
of the main interest. It is given by the following expression
\begin{eqnarray}
E &=& -\frac{e^4}{2}\, \phi^2(0)\,\int\frac{d^4q}{(2\,\pi)^4\,i}\,
\frac{1}{q^4}\,\Bigl[T^{\mu\nu}-t^{\mu\nu}(M)\Bigr]\,t_{\mu\nu}(m)
\nonumber \\ &=&
-2\,e^4\,\phi^2(0)\,\frac{m}{M}\,\int\frac{d^4q}{(2\,\pi)^4\,i}\,
\frac{(T_2-t_2)(q^2-\nu^2)-(T_1-t_1)\,(q^2+2\,\nu^2)}
{q^4\,(q^4-4\,m^2\nu^2)} \label{7}\,,
\end{eqnarray}
where
\begin{eqnarray}
T^{\mu\nu} &=& -i\,\int d^4q\,e^{i\,q\,(x-x')}\, 
\langle P|T\,j^{\mu}(x)\,j^{\nu}(x')|P\rangle \nonumber \\ &=&
-\biggl(g^{\mu\nu}-\frac{q^\mu\,q^\nu}{q^2}\biggr)\,\frac{T_1}{M}+
\biggl(t^\mu-\frac{\nu}{q^2}\,q^\mu\biggr)\,
\biggl(t^\nu-\frac{\nu}{q^2}\,q^\nu\biggr)\,\frac{T_2}{M}\,, \label{8} 
\end{eqnarray}
$t=(1,0,0,0)$, $P=M t$ is a proton momentum at rest and $\nu = q^0$.
For a point--like proton $T^{\mu\nu} \equiv t^{\mu\nu}(M)$ and
\begin{eqnarray}
t^{\mu\nu}(M) &=& {\rm Tr}\biggl[\gamma^\mu\frac{1}{\not\!p-M}\,
\gamma^\nu\,\frac{\gamma^0+I}{4}\biggr]+(q\rightarrow -q)\,, \\
t_1 &=& -\frac{4\,M^2\,\nu^2}{q^4-4\,M^2\,\nu^2}\,, \\
t_2 &=& \frac{4\,M^2\,q^2}{q^4-4\,M^2\,\nu^2}\,.
\end{eqnarray}
Since the amplitude $T^{\mu\nu}$ does not have any singularities at
small $q^2$ the following holds:
\begin{eqnarray}
T_2 &=& O(q^2)\,, \label{12}\\
T_1+\frac{\nu^2}{q^2}\,T_2 &=& O(q^2)\,. \label{13}
\end{eqnarray}
The off--shell (spin averaged) forward
Compton amplitude $T^{\mu\nu}$ of the proton
is not directly measured. However, it could be expressed
in terms of its imaginary part, through the dispersion relations.
Before using them, one notices that the integral in Eq. (\ref{7}) 
is infrared divergent.
It requires subtraction of the leading finite size term, which has already
been accounted for. It also requires an additional subtraction of 
the  proton self--energy term.
Since it changes the analytic behavior of $T_1$ and $T_2$ at small $p^2-M^2$, 
see Eq. (\ref{18}), this proton self--energy  could be only 
partially accounted for in the proton formfactors, 
as a contribution to anomalous magnetic moment or the charge radius.
Since, in our opinion
this problem is not generally known, we repeat here the analysis
from our former paper on radiative recoil corrections and correct
some minor missprints.

If we assume a point-like proton the contribution of the proton self--energy
to the Lamb shift of S-states is \cite{sy} 
\begin{equation}
E = \frac{\alpha^5\,\mu^3}{\pi\,n^3\,M^2}\,\left[\left(
\frac{10}{9}+\frac{4}{3}\,\ln\frac{M}{\mu\,\alpha^2} \right)
-\frac{4}{3}\, \ln k_0(n)\right]\,. \label{14}
\end{equation}
For a {\em true} proton there is a finite size correction 
as given by Eq. (\ref{4}).
The problem is that the proton self-energy is modified by
and modifies as well, the finite size effect. 
Therefore some corrections might be  counted twice. 
To incorporate the  correction (\ref{14}) unambiguously 
we must precisely specify what is the nuclear mean square charge radius.
Its usual definition through the Sachs formfactor 
\begin{equation}
\frac{\langle r^2\rangle}{6} = \frac{\partial G_E({q}^2)}
{\partial ({q}^2)}\biggr|_{{q}^2 = 0}
\end{equation}
is not correct at our precision level, 
because the radiative correction to $G_E$
is infrared divergent or depend on spurious photon mass. 
Following \cite{krp2} we propose thus a different definition using
the forward  scattering amplitude described by $T^{\mu\nu}$,
or more precisely by its longitudinal component $T_L$
defined by
\begin{equation}
T_L \equiv \biggl(1-\frac{\nu^2}{q^2}\biggr)T_2-T_1 \,.
\end{equation} 
For our purpose we consider a nonrelativistic limit 
$\nu \sim { q}^2 $ and
$ p^2-M^2 = (P+q)^2-M^2 \approx 0$ of $T_L$.
For a point-like particle without radiative corrections $T_L$ is
\begin{equation}
T_L \approx \,M\,\mbox{\rm Tr}\left[
\gamma^0\,\frac{1}{\not\!p -M}\gamma^0\,\frac{(\gamma^0+I)}{4}\right]+
(q\rightarrow -q)
\approx \frac{2\,M^2}{p^2-M^2}+(q\rightarrow -q)\,,
\end{equation}
where $p=P+q$. With a finite size particle 
\begin{equation}
\gamma^\mu \rightarrow \Gamma^\mu = 
\gamma^\mu\,F_1 + i\frac{\sigma^{\mu\nu}}{2M}\,q_\nu\,F_2\,,
\label{16}
\end{equation}
$T_L$ acquires a correction
\begin{equation}
\Delta T_L \approx \frac{2\,M^2}{p^2-M^2}[G_E^2(q^2)-1] 
+(q\rightarrow -q) \approx
\frac{2\,M^2}{p^2-M^2}\,q^2\,\frac{\langle r^2\rangle_{bar}}{3}+(q\rightarrow -q)\,,
\end{equation}
where $G_E = F_1 + \frac{q^2}{4\,M^2}\,F_2$ is an electric formfactor.
The radiative corrections for a point-like particle \cite{krp2}
in the nonrelativistic limit  are 
\begin{equation}
\Delta T_L = \frac{\alpha}{\pi}\,\frac{q^2}{p^2-M^2}\,
\left(\frac{10}{9}+\frac{4}{3}\,\ln\frac{M^2}{M^2-p^2}\right)
+(q\rightarrow -q)\,. \label{18}
\end{equation}
We define $\langle r^2 \rangle$ by the following equation
that describes the low-energy behavior of  the
forward scattering amplitude
\begin{equation}
T_L -t_L \approx  \frac{q^2}{p^2-M^2}\left(
\frac{4\,\alpha}{3\,\pi}\,\ln\frac{M^2}{M^2-p^2}+
\frac{2}{3}\,M^2\,\langle r^2\rangle \right)+(q\rightarrow -q)\,.
\label{19}
\end{equation}
We expect that for any nucleus the logarithmic term above will be the same,
since it is only related to the fact that nucleus has a charge, 
and does not depend on other details like the spin. 
The associated correction to the energy for S-states has the form
\begin{equation}
\Delta E = \frac{2}{3\,n^3}\,\alpha^4\,\mu^3\langle r^2\rangle + 
\frac{4\,\alpha^5}{3\,\pi\,n^3}\,\frac{\mu^3}{M^2}\,
\left[\ln\left(\frac{M}{\mu\,\alpha^2}\right)-\ln k_0(n)\right]\,. \label{74}
\end{equation}
The small second term in the above Eq. gives $-0.0099$ meV for 2S state in $\mu$H.
For P-states the proton self--energy contributes only through Bethe log.
The correction to
anomalous magnetic moment is already taken into account
in the calculation of relativistic effects as given by the Breit Hamiltonian. 

We now return to the forward scattering amplitude
and its associated correction to energy of $\mu$H.
We can neglects here QED effects on the structure functions,
since they are $\alpha$ times smaller.
Otherwise dispersion relations for $T_1$ and $T_2$
as in Eq. (\ref{29},\ref{30}) will not be correct.
After neglection of QED effects one can expect a separated pole
at $2M\nu=q^2$, that is due to the elastic contribution. It is
obtained  with the help of the proton elastic formfactors $F_1$ and $F_2$,
see Eq. (\ref{16}).
One derives for $T_1$ and $T_2$ the following expressions
\begin{eqnarray}
T_1^B &=&-\frac{1}{(q^4-4\,M^2\,\nu^2)}\,
(4\,M^2\nu^2F_1^2 + 2\,q^4\,F_1\,F_2+q^4\,F_2^2) \\
T_2^B &=& \frac{1}{(q^4-4\,M^2\,\nu^2)}\,
(4\,M^2\,q^2\,F_1^2-q^4\,F_2^2)\,.
\end{eqnarray}
Any contribution in the two photon exchange
beyond this second order Born term
is considered to be due to the proton polarizability.   
The correction to energy as given by Eq. (\ref{7}) shows infrared singularity,
which is due to the finite size effect, already accounted for at the lower
order in $\alpha$. One first subtracts this singularity and then integrate
over $q$. With the commonly used dipole parametrization of proton
formfactors and $\Lambda^2 = 0.71$ GeV$^2$ one obtains $\Delta E = 0.018$ meV.
However, as pointed out in \cite{hau}
this result is not correct, because 
this dipole parametrization corresponds to the proton radius $r= 0.81$ fm,
which differs significantly from the  value $r= 0.862$ fm.
Instead, we use the parametrization by Simon {\em et al.} \cite{rp}
with the result
\begin{equation}
\Delta E = 0.0232(15)\,{\rm meV} = r^3\,0.0281\,. \label{23}
\end{equation}
This  parametrization in \cite{rp} 
is not correct for $ Q^2 \equiv -q^2 \approx 1$ GeV
or greater since for large $Q^2$ it behaves like $Q^{-2}$ not $Q^{-4}$, 
however it does not influence the result at the presented precision level.
This $r^3$ dependence in Eq. (\ref{23}) is only an approximate dependence.
This dependence become exact in the large nucleus mass limit with  
the dipole parametrization of proton formfactors.
The result $0.0232(15)$ was obtained assuming $r=0.862(12)$ fm.
If the proton radius $r$ from  $\mu$H measurement will 
be significantly different,
this result should be adjusted with the improved parametrization
of proton formfactors. 

Let us assume now, that the second order Born contribution 
is subtracted from $T_1$ and $T_2$.
The proton polarizability correction $E_{POL}$ as given by Eq. (\ref{7})
is split into two parts
\begin{eqnarray}
E_{POL} &=& E_1+E_2\,,\\
E_1 &=& -2\,e^4\,\phi^2(0)\,\frac{m}{M}\,\int\frac{d^4 q}{(2\,\pi)^4\,i}\,
\frac{q^4+2\,\nu^4}{q^6(q^4-4\,m^2\,\nu^2)}\,T_2(\nu,q^2)\,,\label{27} \\
E_2 &=& 2\,e^4\,\phi^2(0)\,\frac{m}{M}\,\int\frac{d^4 q}{(2\,\pi)^4\,i}\,
\frac{q^2+2\,\nu^2}{q^4\,(q^4-4\,m^2\,\nu^2)}\,\biggl(
T_1(\nu,q^2)+\frac{\nu^2}{q^2}\,T_2(\nu,q^2)\biggr)\,. \label{28}
\end{eqnarray}
The first dominant part could be well calculated, while the calculation
of the smaller second part will require additional assumptions.
This second part was usually neglected in studies of polarizability effect
in composite nuclei, see for example \cite{be}.
The  imaginary part of $T_i$  in variable $\nu$
at fixed $q^2$ is obtained from the inclusive cross section
$\gamma^*+p\rightarrow X$. The real part could be restored from the 
imaginary one using the following dispersion relations:
\begin{eqnarray}
T_2(\nu,q^2) &=& -\int_{\nu_{th}^2}^\infty\,d\nu'^2\,
\frac{W_2(\nu',q^2)}{\nu'^2-\nu^2}\,, \label{29}\\
T_1(\nu,q^2) &=& T_1(0,q^2)
-\nu^2\,\int_{\nu_{th}^2}^\infty\,\frac{d\nu'^2}{\nu'^2}\,
\frac{W_1(\nu',q^2)}{\nu'^2-\nu^2}\,,
\label{30}
\end{eqnarray}
where $\nu_{th}$ is the threshold value of $\nu$  for the production
of $\pi$ mesons.
This formulas are obtained from \cite{bj}, by multiplying right hand side 
by $-1/2$, due to the different definition of $T_i$ in our work.
While $W_2$ for proton behaves like $1/\nu$ to large $\nu$, 
$W_1$ goes like $\nu$, and thus the dispersion relation in (\ref{30})
contains a subtraction at $\nu=0$. We have assumed here, that $T_2$
has similar asymptotic behavior as $W_2$. i.e. vanishes 
in the large $\nu$ limit. However, there is no too much information
on $T_1(0, q^2)$. One recognizes that in the limit of small $q^2$
it is given by the magnetic polarizability $\beta_M$
\begin{equation}
\lim_{q^2\rightarrow0} \frac{T_1(0,q^2)}{q^2} = \frac{M}{\alpha} \,\beta_M \,,
\label{31}
\end{equation}
which amounts to $\beta_M = 1.56(57)\cdot 10^{-4}$ fm$^3$, see \cite{beta}.
Since there is no experimental data,
we assume here, that the $q$-dependence is governed by
the square of the elastic formfactor, namely
\begin{equation}
\beta_{M}(Q^2) = \beta_{M}\,\frac{\Lambda^8}{(\Lambda^2+Q^2)^4}
\end{equation}
with $\Lambda^2=0.71$ GeV$^2$.
Using equations (\ref{27}-\ref{31})
one derives the following expressions for $E_1$ and $E_2$.
\begin{eqnarray}
E_1&=& -\alpha^2\,\frac{\phi^2(0)}{m\,M}\,
\int_{\nu_{th}^2}^\infty d\nu^2\,\int_0^\infty \frac{dt}{t^2}\,
g(\nu,t)\,W_2(\nu,-t)\,, \label{33}\\ \nonumber \\
E_2&=& -\alpha^2\,\frac{\phi^2(0)}{m\,M}\,
\int_{\nu_{th}^2}^\infty d\nu^2\,\int_0^\infty \frac{dt}{t^2}\,
f(\nu,t)\,W_2(\nu,-t)\,\frac{1}{1+R}\,\left(1-R\,\frac{\nu^2}{t}\right)
\nonumber \\ &&
+\alpha^2\,\frac{\phi^2(0)}{m\,M}\,
\int_0^\infty dt \,h(t)\,\frac{M}{\alpha}\,\beta_M(t)\,, \label{34}
\end{eqnarray}
where
\begin{eqnarray}
g(\nu,t) &=& \frac{4}{\pi}\,\int_0^1 dx\,\sqrt{1-x^2}\,\,
\frac{(1+2\,x^4)}{2\,(x^2+\frac{t}{4\,m^2})(x^2+\frac{\nu^2}{t})}\,,\\
\nonumber \\
f(\nu,t) &=& \frac{t}{2\,\nu^2}\,\frac{4}{\pi}\,\int_0^1 dx\,\sqrt{1-x^2}\,
\frac{x^2(1+2\,x^2)}{(x^2+\frac{t}{4\,m^2})(x^2+\frac{\nu^2}{t})}\,,\\
\nonumber \\
h(t) &=& \left[1+\left(1-\frac{t}{2\,m^2}\right)\,
\left(\sqrt{\frac{4\,m^2}{t}+1}-1\right)\right]\,, 
\end{eqnarray}
and $R$ is equal to $ R= {W_L(\nu,q^2)}/{W_1(\nu,q^2)}=\sigma_L/\sigma_T$,
the ratio of longitudinal to transverse cross sections, and thus
could be measured. A recent work \cite{abe} by E143 collaborations
brings the most recent parametrization of $R$ 
in the momentum range $Q^2\geq 1$ GeV. What we find, 
is a weak dependence of $R$ on $\nu$ for small $Q^2$, of order few  GeV,
see for example Figure 3 of this work. Moreover $R$ is between
$0.3 - 0.4$. However there is no precise data on the most important 
region for us of $Q^2<1$ GeV. One knows from conditions (\ref{12},\ref{13}), 
that $R \sim Q^2$ at small $Q^2$ and constant $\nu$. The only known
paper by Brasse {\em et. al} \cite{bras} brings comments on $R$, that it 
is of order 10\% at the resonance region. Since, what we find later on,
$E_2$ is small and we do not have precise data on $R$ at the resonance
region we simply neglect $R$ in the expression for $E_2$ in (\ref{34}). 
We expect, this approximation will not alter significantly
the final result for the polarizability correction.
Various parametrization of structure functions  $W_2$
were presented in the literature. We have chosen that,
which  match the $q^2=0$ limit known as a photoproduction
and are correct for region of  $\nu$ and $Q^2$ of order few GeV.
We use the parametrization from \cite{bras} for the resonance region.
It is given by a compact formula and a long table of numbers, 
see Table 1 of \cite{bras}. For $\nu$ close to the threshold
for $\gamma^* + p^+ \rightarrow \pi^+ + n$ process, we assume a behavior
$\sqrt{W-W_{th}}$ and match the value at $W=1.110$ GeV, as given in \cite{bras}.
The photoproduction of $\pi^0$ could be neglected close to this threshold value.
Out of the resonance region, $W>1.990$ GeV, we use
an ALLM97 parametrization with the recent corrections \cite{allm}.
With these parametrizations
integrals in (\ref{33},\ref{34}) are performed numerically with results
\begin{equation}
E_{POL} = E_1+E_2 = -\alpha^2\,\frac{\phi^2(0)}{m\,M}\, (1.82+0.12-0.26)
                  =-\alpha^2\,\frac{\phi^2(0)}{m\,M} 1.68(33)\,,
\end{equation}  
where the uncertainty of 0.33 forms 20\% of final result and is an estimate
due to approximations and assumptions performed
during this derivation, namely: unknown $R$, unknown $Q^2$
dependence of $\beta_M$ and inaccuracy of ALLM97 parametrization
at low $Q^2$ and $\nu$. The contribution to the (2P-2S)
splitting of $n=2$ states is
\begin{equation}
E_{POL} = 0.012(2)\,{\rm meV}\,.
\end{equation}
It is slightly lower than the result obtained by Rosenfelder in \cite{ros}.
We think it is due to approximate treatment of the $Q$-dependence in his work.
Other contributions to the Lamb shift in muonic hydrogen are
presented in the Table I. It is an improved version of the former 
Table I in \cite{krp1}. All QED corrections are recalculated with a better precision.
The final result for the Lamb shift differs from that in \cite{krp1}
by more than the error estimate there. It is due to the mistake in the sum in 
Table I in \cite{krp1} 
and due to the inclusion of new terms: proton polarizability,
third order electron v.p. and the hadronic v.p. 
It should be noted here that the same effects should be included
in hydrogenic Lamb shift, when having a new value of the proton charge radius.
It was not necessary so far since in the  electron scattering
measurements of proton radius, these effects have not been excluded.
The uncertainty in the final result has two main sources:
the proton polarizability and the estimate of higher order QED effects.
While the first one is pretty difficult to improve, the second source
of uncertainty, higher order QED corrections could be well calculated.
By these higher corrections we mean diagrams presented on Figure 1.
Diagrams in the second raw, have been calculated in
an approximate way in \cite{krp1} and are named in the Table I as a
muon self--energy with electron v.p. Some further theoretical work
is necessary for obtaining the improved value of the hyperfine splitting of S-levels,
since the proton polarizability effect is also present there. 
In summary, having now precise theoretical predictions, the measurement of
the 2P-2S transition frequency in the muonic hydrogen will lead to the
improved value of the proton charge radius.

\section*{Acknowledgments}
I wish to thank M. Krawczyk, E. Rondio for information on
proton structure functions and F. Kottmann
for pointing out a mistake in the sum of all contributions
to the $\mu$H Lamb shift in \cite{krp1}. This  work was  supported by
Polish Committee for Scientific Research under contract No. 2 P03B 024 11

\begin{minipage}[c]{4.5in}
\begin{table}
\begin{center}
\begin{tabular}{|l|r|}  
correction & value in meV \\ \hline 
leading order e.v.p. &                  205.0074 \\ 
rel. corr. to e.v.p . &                 0.0594   \\ 
double e.v.p. &                         0.1509   \\ 
two-loop e.v.p. &                       1.5079   \\ 
three-loop e.v.p. &              0.0076   \\ 
muon self-energy + muon v.p. &     $-0.6677$   \\ 
muon self-energy with e.v.p. & $-0.006(1)$  \\ 
recoil of order $\alpha^4$  &         0.0575   \\ 
recoil of order $\alpha^5$  &        $-0.0450$   \\ 
recoil of order $\alpha^6$  &        $ 0.0003$   \\ 
proton self energy &                 $-0.0099$   \\ 
leading finite size of order $\alpha^4$
 &   $-r^2\, 5.1974 =  -3.862(108)$            \\ 
finite size of order $\alpha^5$ & $r^3\,0.0363 =   0.0232(15)$   \\ 
finite size of order $\alpha^6$ & $-0.0009(3)$   \\ 
e.v.p. with finite size & $-r^2\,0.0281 = -0.0209(6)$ \\ 
hadronic v.p. &                      0.0113(3) \\ 
proton polarizability &              0.012(2)  \\  
estimate for uncalculated terms & (0.002)        \\ \hline
sum of corrections to Lamb & $206.085(3) - r^2\, 5.2255 + r^3\,0.0363$ \\ 
shift in $\mu$H with $r=0.862(12)$& $ = 202.225(108)$  \\ 
\end{tabular}
\end{center}
\caption{Summary of results for corrections to the Lamb shift 
in muonic hydrogen, e.v.p. denotes the electron vacuum polarization.
QED corrections are calculated according to Ref. \protect \cite{jpb}
and \protect \cite{krp1}.
 \label{table1}}
\end{table}
\end{minipage}

\begin{figure}
\centerline{\psfig{figure=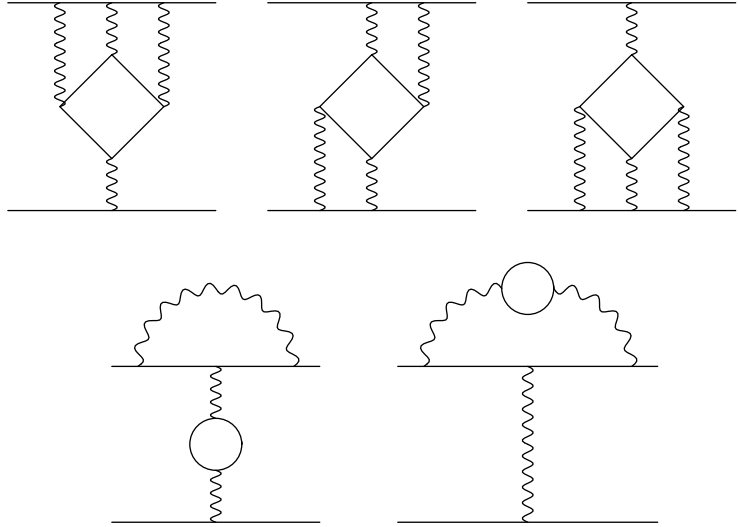,width=4in}}
 \caption{Higher order diagrams contributing to muonic hydrogen Lamb shift.
          The upper  and lower horizontal lines denote the muon and the
          proton respectively. A closed line denotes the electron loop.  }
\label{fig}
\end{figure}
\end{document}